\documentclass[preprint2,times,tighten]{aastex6}

\usepackage{color}
\usepackage{enumitem}
\usepackage{hyperref}
\usepackage{verbatim}
\usepackage{amsmath,amstext,mathrsfs}
\usepackage[all]{hypcap} 
\usepackage{afterpage}
\usepackage{graphicx}

\begin{document}
\title{Studying the local magnetic field and anisotropy of magnetic turbulence by synchrotron polarization derivative }

\author{Jian-Fu Zhang\altaffilmark{1,2}, Ke Hu\altaffilmark{1}, Jungyeon Cho\altaffilmark{3} and Alex Lazarian \altaffilmark{4}} 
\email{hooke@xtu.edu.cn (KH); jfzhang@xtu.edu.cn (JFZ); jcho@cnu.ac.kr (JC); alazarian@facstaff.wisc.edu (AL)}
\altaffiltext{1}{Department of Physics, Xiangtan University, Xiangtan, Hunan 411105, China}
\altaffiltext{2}{Guizhou Provincial Key Laboratory of Radio Astronomy and Data Processing, Guiyang, Guizhou 550025, China}
\altaffiltext{3}{Department of Astronomy and Space Science, Chungnam National University, Daejeon, Republic of Korea}
\altaffiltext{4}{Astronomy Department, University of Wisconsin, Madison, WI 53711, USA}
\begin{abstract}
Due to the inevitable accumulation of the observed information in the direction of the line of sight, it is difficult to measure the local magnetic field of MHD turbulence. However, the correct understanding of the local magnetic field is a prerequisite for reconstructing the Galactic 3D magnetic field. We study how to reveal the local magnetic field direction and the eddy anisotropy on the basis of the statistics of synchrotron polarization derivative with respect to the squared wavelength $dP/d\lambda^2$. In the low frequency and strong Faraday rotation regime, we implement numerical simulations in the combination of multiple statistic techniques, such as structure function, quadrupole ratio modulus, spectral correlation function, correlation function anisotropy and spatial gradient techniques. We find that (1) statistic analysis of $dP/d\lambda^2$ indeed reveals the anisotropy of underlying MHD turbulence, the degree of which increases with the increase of the radiation frequency; (2) the synergy of both correlation function anisotropy and gradient calculation of $dP/d\lambda^2$ enables the measurement of the local magnetic field direction. 
\end{abstract}


\section{Introduction}\label{intro}
Turbulence is a ubiquitous phenomenon in astrophysical plasmas. Due to the electrical conductivity of plasma astrophysics, turbulent motions of astrophysical fluids are accompanied by magnetic field fluctuations (\citealt{Biskamp03}), resulting in the production of magnetohydrodynamic (MHD) turbulence. The most significant observational evidence for turbulence is from the spectral distribution of electron density fluctuations in the Milky Way (\citealt{Armstrong95,Chepurnov10}). The extra evidence comes from measurements of velocity fluctuations by using Doppler shifted spectral lines (\citealt{L09}, for a review) and magnetic field measurements by synchrotron fluctuations (\citealt{Cho10,Gaensler11,Burkhart12}). Actually, MHD turbulence plays a major role in the key astrophysical processes, such as star formation (\citealt{2004RvMP...76..125M,2007ARA&A..45..565M}), propagation and acceleration of cosmic rays (\citealt{Yan08}), heat conduction (\citealt{2001ApJ...562L.129N}), and turbulent magnetic reconnection (\citealt{Lazarian99}, hereafter LV99). The importance of studying MHD turbulence lies in that 
we can understand the processes of astrophysics mentioned above and improve theoretical development as well.

In recent three decades, traditional theoretical and experimental methods in the study of MHD turbulence have been complemented by numerical simulations providing many valuable results, such as the confirmation of scale-dependent anisotropy (\citealt{ChoV00}, hereafter CV00) and the compressibility of MHD turbulence (\citealt{Cho02PRL,Kowal10}). Usually, the Reynolds numbers $R_{\rm e}= LV/\mu$, used as characterizing the level of turbulence, are extremely large in the magnetized astrophysical plasmas, with the large astrophysical scale $L$ and velocity $V$ but the finite value of kinematic viscosity and magnetic diffusivity $\mu$ determined by microphysics. An example is that the order of magnitude $R_{\rm e}=10^{10}$ (or larger) are common for the interstellar medium (ISM). However, the present direct numerical simulations are limited with $R_{\rm e}\sim10^5$, leaving a large gap between the numerical simulation setting and real astrophysical environment. Therefore, it is necessary to develop and use new observation-oriented methods for measuring the properties of ISM, removing the Cosmic Microwave Background (CMB) foreground, exploring the propagation of cosmic rays and predicting the ratio of star formation. 

Many astrophysical phenomena suggest that relativistic electrons are widespread in cosmos, where interactions of the electrons with the fluctuating magnetic fields lead to synchrotron emission fluctuations carrying important information of magnetization turbulence. By studying synchrotron fluctuations can one obtain basic insights into the magnetic fields. We would like to emphasize how relativistic electrons are accelerated is still an open issue. According to the classical shock acceleration theory, the single spectral index 2 of the electrons was usually assumed in a number of earlier papers. Hence, synchrotron intensity fluctuations are used to extract the spectrum and anisotropies of underlying magnetic turbulence (e.g., \citealt{Getm59,L90}), and polarization fluctuations to measure magnetic field helicity (\citealt{Wael09,Junk11}). Adopting an analytical description, Lazarian \& Pogosyan (2012, hereafter LP12) derived in detail the correlation of synchrotron fluctuations for an arbitrary spectral index of relativistic electrons and predicted the synchrotron intensity fluctuations anisotropic with significant correlation along the direction of the magnetic field. One of their main findings is that the anisotropy is dominated by the quadrupole moment sensitive to the compressibility of the underlying turbulence, which was confirmed in \cite{Herron16} using numerical simulations. Recently, the quadrupole ratio modulus has been applied to the study of synchrotron polarization intensity anisotropy (\citealt{Lee19,Wang20}).

The polarized synchrotron radiation together with the Faraday rotation measure, known as the Faraday rotation synthesis (\citealt{Burn66,Brentjens05}), can provide valuable information on the plane-of-sky geometry of the magnetic field for external galaxies. The difficulty in studying the line-of-sight magnetic field component using this method is obvious (\citealt{Ferriere16}) for the sign of the Faraday rotation measure varies due to the direction changes of the line-of-sight magnetic field component. To break the limit of this technique, Lazarian \& Pogosyan (2016, hereafter LP16) proposed a series of theoretical predictions of synchrotron polarization fluctuations to reveal the underlying magnetic turbulence, which were confirmed in the consequent numerical simulations (\citealt{Zhang16,Zhang18,Lee16}) and applied to observations (\citealt{Xu16,GuoMW17}; see \citealt{BL19} for a recent book). 

The scale-dependent anisotropy of eddies is a significant characteristics of the theory of MHD turbulence (\citealt{Gold95}, henceforth GS95). That the degree of the scale-dependent anisotropy of eddies enlarges as the eddy size gets smaller (GS95) is established on the local magnetic field direction (LV99; CV00) which was supplemented to the MHD turbulence theory in the later studies (LV99; CV00). We would like to emphasize that the maintaining of the scale-dependent anisotropy is not in the observer (projected mean magnetic field) reference frame, where the largest eddies endow MHD turbulence with the scale-independent anisotropy (CV00).

It is in the reference system of the mean magnetic field that theoretical descriptions of both synchrotron intensity (LP12) and polarization intensity fluctuations (LP16) are formulated along the line-of-sight integral. Adopting the eddy description of MHD turbulence (LV99; \citealt{ChoLV02}), the gradients of both velocities and magnetic field are expected to be perpendicular to the local magnetic field direction, which motivated the development of gradient techniques of the velocity and synchrotron radiation (\citealt{GCL17,Lazarian17}). The synchrotron gradient techniques, including synchrotron intensity (\citealt{Lazarian17}), polarization intensity and its derivative with respect to the squared wavelength (\citealt{LY18}, henceforth LY18), were suggested to trace directions of projected mean magnetic fields. By promoting polarization gradients to various advanced diagnostic quantities derived in \cite{Herron18a,Herron18b}, \cite{Zhang19a,Zhang19b} found that synchrotron gradient techniques take a significant advantage over the traditional polarization method in tracing projected magnetic fields, and consistent measurements are made for the Galactic magnetic field directions by means of the multifarious advanced diagnostic gradients.

With the 3D data cubes of magnetic fields from numerical simulations, CV00 has indeed achieved the scale-dependent anisotropy in the local frame of the eddies, using the second-order structure function. However, when it comes to observations, it seems to be prohibitive to obtain the local magnetic field due to the inevitable accumulation of the observed information in the direction of the line of sight. The reproduction of the local magnetic fields is essential to reconstruct the Galactic 3D magnetic field structure. As in our previous work, the synchrotron polarization derivative with regard to the squared wavelength $dP/d\lambda^2$ was predicted analytically by LP16 and confirmed numerically by \cite{Zhang18} to be sensitive to Faraday rotation. Using statistic analysis of $dP/d\lambda^2$, this paper studies how to obtain the anisotropy of the small-scale eddies and trace the direction of the local magnetic field in the case of the very low-frequency strong Faraday rotation. 

The paper is developed as follows. Section 2 gives brief descriptions on the theory of MHD turbulence, synchrotron radiative processes and polarization derivative with respect to the squared wavelength, following the description of multiple statistic techniques in Section 3. Section 4 provides the description of the generation and anisotropy analysis of simulation data cubes, with numerical results presented in Section 5.  Discussion and summary are made in Sections 6 and 7, respectively. 

\section{Theoretical fundamentals}\label{DTFsect}
\subsection{Scale-dependent Anisotropy of MHD Turbulence}
Since the hydrodynamic type motions of eddies that mix magnetized fluid give rise to Alfvénic perturbations for the magnetic field, MHD turbulence was usually considered as the dynamical processes of the turbulent cascade dominated by the Alfvén wave interactions. The Alfvén Mach number, describing the strength of magnetic turbulence, is defined as $M_A=V_{L}/V_A$, where $V_{L}$ is turbulent injection velocity at the scale $L_{\rm in}$ and $V_{\rm A}= {\langle B \rangle}/\sqrt{4\pi \rho}$ is the Alfvén velocity. Correspondingly, the largest turbulent eddies would also reflect the Alfvén Mach number at the injection scale $L_{\rm in}$, i.e., $M_A=B_L/{\langle B \rangle}$ with $B_{L}$ being the perturbation of the magnetic field $B$ at the injection scale $L_{\rm in}$. Then, the relative perturbations of magnetic fields and eddy velocities are related in a symmetrical way (\citealt{Lazarian20})
\begin{equation}
\frac{\delta B_l}{{\langle B \rangle}}=\frac{\delta B_l B_{L} }{B_{L} {\langle B \rangle}}=\frac{v_l}{V_{L}}M_{\rm A}=\frac{v_l}{V_{\rm A}},
\end{equation}
where $\delta B_l$ is the fluctuation of the magnetic field $B$ at the scale $l$ of the turbulent fluid and $v_l$ is the corresponding velocity fluctuation. 

The modern theory of MHD turbulence (GS95) can be understood as a collection of anisotropic eddies aligning with the direction of local magnetic fields around them, that is, the motions of the eddies with parallel scales $l_\parallel$ and perpendicular scales $l_\perp$ are elongated along the direction of the local magnetic field. Assuming that the motion perpendicular to the direction of the local magnetic field has a Kolmogorov scaling of $v_l\sim l_\perp^{1/3}$, the anisotropic relation of eddy scales can be written as 
\begin{equation}
l_{\|}\propto l_{\perp}^{2/3}. \label{eq:aniso}
\end{equation}
This scaling was originally predicted in the case of strong trans-Alfv{\'e}nic ($M_{\rm A}\sim 1$) incompressible turbulence, with the critical balance condition of $l_\perp^{-1}v_\perp \sim l_\parallel^{-1} V_{\rm A}$. Later, the GS95 anisotropic scaling (see Equation (\ref{eq:aniso})) was generalized to both $M_{\rm A}>1$ and $M_{\rm A}<1$ cases (LV99, see also \citealt{Lazarian06}). The former is called super-Alfv{\'e}nic turbulence regime with $V_{L}> {V}_{{\rm{A}}}$. In the limiting case of $V_{L}\gg V_{\rm A}$, the turbulence presents an essentially hydrodynamic Kolmogorov behavior, i.e., $v_l=V_{{L}}(l/{L_{\rm in}})^{1/3}$, as the weak magnetic field with lower magnetic energy than kinetic energy has a marginal effect on turbulent dynamics. When $v_{l}=V_{\rm A}$, the hydrodynamic-like behavior of the turbulence cascade would transition into strong turbulence regime at the scale $l_{\rm A}={L_{\rm in}}M_{\rm A}^{-3}$. In the range from $l_{\rm A}$ to the dissipation scale $l_{\rm dis}$, the GS95 scaling can be maintained.  

The latter, $V_{{L}}< {V}_{{\rm{A}}}$, shows weak turbulence ranging from ${L_{\rm in}}$ to the transition scale $l_{\rm B}={L_{\rm in}}M_{\rm A}^2$, in which the magnetic field fluctuations are quasi-2D and perpendicular to the direction of magnetic fields (LV99, \citealt{Galtier00}). At scales less than $l_{\rm B}$, the relation between the eddy major axis $l_\|$ and its minor axis $l_\bot$ is present again, i.e.,
\begin{equation}
l_{\|}\approx {L_{\rm in}}( l_\perp/L_{\rm in})^{2/3} M_{\rm A}^{-4/3}. \label{eq:aniso1}
\end{equation}
Setting $M_A=1$, we recapture the original GS95 relation (see Equation (\ref{eq:aniso})). 

In compressible MHD turbulence, the compressible turbulent motions are associated with slow and fast modes (GS95; \citealt{Lithwick01}). Numerical simulations (\citealt{Cho02PRL}) confirmed that Alfvénic and slow modes have the same anisotropy ($l_{\|}\propto l_{\perp}^{2/3}$), while fast modes present isotropic cascade ($l_{\|}\propto l_{\perp}$).  

\subsection{Synchrotron Polarization Radiative Process}\label{SynRP}
The interactions of relativistic electrons with turbulent magnetic fields emit synchrotron radiation which reveals the information from magnetic field fluctuations. For simplicity, we consider a homogeneous and isotropic distribution of the electrons with a power-law form of $N(\gamma) \propto \gamma^{2\alpha-1}$, where $\gamma$ is the electron energy and $\alpha$ is a spectral index of the electrons\footnote{In our simulations, a power-law energy distribution of the isotropic relativistic electrons is assumed to produce the synchrotron polarization emission. However, the interested reader should notice the following possible cases happening in the astrophysical environments. First, a hybrid thermal and non-thermal electron distribution (\citealt{Mao18b}) occurs due to the turbulence heating and acceleration to electrons, the temperature of which is associated with the electron energy distribution that deviates from a Gaussian- or blackbody-like distribution (\citealt{Mao18a}). In this regard, we predict that the anisotropy of polarization radiation should be related to the temperature of the electrons accelerated. Second, given that the relativistic electron distribution is anisotropic, such as in the region of strong shocks accompanied by turbulence (called jitter radiation: application to gamma-ray burst (\citealt{Mao13}) and X-ray binary (\citealt{Zhang17})), the anisotropic synchrotron polarization radiation will enhance the polarization degree of synchrotron radiation, leading to more significant anisotropy. This would help us improve the anisotropy analysis of polarization techniques and the ability of magnetic field tracing.}. Thus, the synchrotron emission intensity as a function of radiative frequency $\nu$ can be written as (\citealt{Ginzburg65})
\begin{eqnarray}
I(\nu)\propto &&  \frac{e^3}{4\pi m_{\rm e} c^2}  \frac{\sqrt 3}{2-2\alpha}
\Gamma \left(\frac{2-6\alpha}{12}\right)\Gamma \left(\frac{22-6\alpha}{12}\right)
\nonumber \\
&&   \left(\frac{3e}{2\pi m_{\rm e}^3 c^5}\right)^{-\alpha} 
\int_0^L B_{\perp}^{1-\alpha} \nu^{\alpha}dL, 
\label{eq:I}
\end{eqnarray}
where $\Gamma (x)$ is a gamma function, $B_{\perp}$ the magnetic field component perpendicular to the line of sight, and $L$ an emitting-region size along the line of sight. Other parameters ($e$, $m_{\rm e}$ and $c$) have their usual meanings. With the spectral index of relativistic electrons, we would gain the fraction polarization degree $p=\frac{3-3\alpha}{5-3\alpha}$ and the linearly polarized intensity $P=pI$. The observable Stokes parameters $Q$ and $U$ are related to the polarized intensity by the polarization angle $\psi_0$, i.e., $Q=P\cos 2\psi_0$ and $U=P\sin 2\psi_0$. When it comes to a Faraday rotation effect, the polarization angle is expressed as $\psi=\psi_{0}+\varPhi \lambda^2$ with the wavelength $\lambda$. The Faraday rotation measure $\varPhi$ is given by    
\begin{equation}
\varPhi (\textbf{\textit{X}},z)={e^3\over2\pi m_{\rm e}^2c^4}\int^z_{0}dz'n_{\rm e}(\textbf{\textit{X}},z')B_\|(\textbf{\textit{X}},z'). \label{RM}
\end{equation}
Here, $\textbf{\textit{X}}=(x,y)$ represents a vector in the plane of the sky, $B_{\|}$ the parallel component of the magnetic field, and $n_{\rm e}$ the thermal electron density.

\subsection{Synchrotron Polarization Derivative}\label{SynPD}
With the Stokes parameters $Q$ and $U$, the complex polarization vector $\textbf{\textit{P}}=Q+iU$ is formulated as (LP16) 
\begin{equation}
\textbf{\textit{P}}(\textbf{\textit{X}},\lambda)=\int^{L}_{0}dzP_i(\textbf{\textit{X}},z){e}^{2i\lambda^2\varPhi(\textbf{\textit{X}},z)}, \label{PEq}
\end{equation}
where $P_i(\textbf{\textit{X}},z)$ indicates the intrinsic polarized intensity density treated as wavelength independent; this simplified treatment would not change our numerical results stated below (see also \citealt{Zhang18} for confirmation). 

The condition of de-correlation of the Faraday rotation measure is introduced as 
\begin{equation}
\label{eq:oneradcon}
\lambda^2\varPhi = 0.81\lambda^2 \int_0^{L_{\rm eff}} dz n_{\rm e} B_\| \approx 2\pi,
\end{equation}
where $L_{\rm eff}$ is an effective spatial depth of the Faraday rotation sampling. Therefore, the ratio of the depth sampled by the Faraday rotation to the emitting region size can be written as
\begin{equation}
\label{eq:el}
\frac{L_{\rm eff}}{L} \sim \frac{2\pi}{\lambda^2L} \frac{1}{\phi}, 
\end{equation}
where $\phi= {\rm max}(\sqrt{2} \sigma_\phi,\bar{\phi})$, $\sigma_\phi$ and $\bar{\phi}$ are the root mean square and the mean of the Faraday rotation measure density ($n_{\rm e} B_\|$) fluctuation, respectively. Using Equation (\ref{eq:el}), the strong and weak Faraday rotation would be characterized by $L_{\rm eff}/L<1$  and $L_{\rm eff}/L>1$, respectively. When $L_{\rm eff}/L<1$,  Equation (\ref{PEq}) can be split two parts, in which only the part of $z<L_{\rm eff}$ suffers from the Faraday depolarization while the part of $z>L_{\rm eff}$ cannot contribute to the information of the polarization measure (see Figure 15 in LY18 for an illustration). 

As for a chosen wavelength $\lambda$, the complex polarization vector (see Equation (\ref{PEq})) in an effective Faraday depolarization region could be rewritten as 
\begin{equation}
\begin{aligned}
\label{eq:leffP}
\textbf{\textit{P}}(\textbf{\textit{X}},\lambda) = \int_0^{L_{\rm eff}(\lambda)} dz P_i(\textbf{\textit{X}}, z) e^{2i\lambda^2 \varPhi(\textbf{\textit{X}}, z)}.
\end{aligned}
\end{equation}
Considering two neighboring wavelengths $\lambda_1$ and $\lambda_2$, which respectively correspond to the close spatial positions $L_1$ and $L_2$, we can obtain the difference both $\textbf{\textit{P}}(\textbf{\textit{X}},\lambda_1)$ and $\textbf{\textit{P}}(\textbf{\textit{X}},\lambda_2)$ 
\begin{equation}
\begin{aligned}
\label{eq:diffP}
\Delta \textbf{\textit{P}}(\textbf{\textit{X}}) 
&\approx \int_{L_2(\lambda_2)}^{L_1(\lambda_1)} dz P_i(\textbf{\textit{X}}, z) e^{2i\lambda^2\varPhi(\textbf{\textit{X}}, z)},
\end{aligned}
\end{equation}
which reflects the information of the local magnetic turbulence within $z\in [L_2,L_1]$ region. From an observational point of view, providing the Stokes parameters $Q$ and $U$ at two neighboring frequencies, we can reveal the information of magnetic field fluctuation originating from a locally spatial region. However, it should be noted that there is a non-linear correspondence between $L$ and $\lambda$, as expressed in Equation (\ref{eq:el}). In practice, we would consider the statistics of synchrotron polarization intensity derivative with respect to the squared wavelength
\begin{equation}
\label{eq:SPDGa}
\frac{d|\textbf{\textit{P(\textbf{\textit{X}})}}|}{d\lambda^2} = \frac{ |\textbf{\textit{P}}(\textbf{\textit{X}}, \lambda_2)-\textbf{\textit{P}}(\textbf{\textit{X}}, \lambda_1)|}{\lambda^2_2-\lambda^2_1},
\end{equation}
to reveal the local magnetic field fluctuations.

\section{Statistic Measurement Methods}
\subsection{Correlation and Structure Function Anisotropies}\label{CSFA}
The spatial correlation and structure functions of (any) physical variable $f(\textbf{\textit{X}})$ have been traditionally employed for studying anisotropic properties. The former, i.e., correlation function is written as 
\begin{equation}
{\rm CF}(\textbf{\textit{X}})=\left< f(\textbf{\textit{X}}_{\rm 1})f(\textbf{\textit{X}}_{\rm 2})\right>, \label{CFp}
\end{equation}
where, $\left <...\right>$ denotes an average over the entire volume of interest. As for the latter, the commonly used second-order structure function is given by
\begin{equation}
{\rm SF}(\textbf{\textit{X}})=\left< (f(\textbf{\textit{X}}_{\rm 1})-f(\textbf{\textit{X}}_{\rm 2}))^2\right>=2\left [{\rm CF}(0)-{\rm CF}(\textbf{\textit{X}})\right ]. \label{SFqe}
\end{equation}
It is obvious that the second-order structure function is formally related to the correlation function. Both of them can be used to characterize the eddy anisotropy in MHD turbulence. For the structure function, considering two measurement directions perpendicular to each other, we can define their ratio of individual structure function as 
\begin{equation}
\frac{{\rm SF}_x}{{\rm SF}_y}=
\frac{\langle{\vert f_x(x+\Delta x, y)-f_x(x, y)\vert}^2\rangle}
{\langle{\vert f_y(x, y+\Delta y)-f_y(x, y)\vert}^2\rangle}, \label{eq:SFxSFy}
\end{equation}
which would reflect the eddy anisotropy of the turbulence structure.

As for the correlation function, it is easy to calculate the correlation function map with a periodic boundary condition through the fast Fourier transform. Studying the non-periodic locality of synchrotron polarization derivative map, we use the Hockney method (\citealt{Hock68}) to solve the open-boundary convolution problem, which help us decrease computational complexity (\citealt{Yuen18}). The direction of the major axis of correlation (structure) function contour would determine the eddy major axis reflecting the magnetic field orientation. 

\subsection{Spectral Correlation Function }\label{SpecCF}
The spectral correlation function (SCF) was initially introduced in \cite{Lazarian02} to study the anisotropy of turbulence in channel maps of the position-position-velocity 3D space. Here, we adopt this method to explore the correlation of 3D data cubes (as a function of position-position-frequency $f(x,y,\nu)$). The formula used is given by (e.g., \citealt{Padoan03})
\begin{equation}
{\rm SCF}(\textbf{\textit{R}})=1- \left< \sqrt{\frac{\sum_{\nu}|f(\textbf{\textit{X}}+\textbf{\textit{R}}, \nu)-f(\textbf{\textit{X}}, \nu) |^2 }{\sum_\nu |f(\textbf{\textit{X}}+\textbf{\textit{R}}, \nu)|^2+\sum_\nu |f(\textbf{\textit{X}}, \nu)|^2} }  \right>, \label{eq:scf}
\end{equation}
where $\textbf{\textit{R}}=(\Delta x, \Delta y)$ is a lag vector. The contour of ${\rm SCF}( \textbf{\textit{R}})$ would characterize the spatial scales that the spectral features begin to change. The 2D ${\rm SCF}(\textbf{\textit{R}})$ correlation can be used to produce the 1D spectrum of the correlation vs the lag length $R=\sqrt{\Delta x^2 + \Delta y^2}$ by an azimuthal average.

\subsection{Quadrupole Ratio Modulus  }\label{QRM}
Following LP12, the normalized correlation function (NCF) of (any) physical variable $f(\textbf{\textit{X}})$ can be written as 
\begin{equation}
{\rm NCF}(\textbf{\textit{X}})=\frac
{{\langle f(\textbf{\textit{X}})f(\textbf{\textit{X}}+\Delta \textbf{\textit{X}})\rangle}-{\langle f(\textbf{\textit{X}})\rangle}^2}
{\langle f(\textbf{\textit{X}})^2 \rangle-{\langle f(\textbf{\textit{X}})\rangle}^2},\label{eq:NCF}
\end{equation}
where $\Delta \textbf{\textit{X}}$ is a separation vector between any two spatial points on the plane of the sky. Similarly, the normalized structure function is expressed as  
\begin{equation}
{\rm NSF}(\textbf{\textit{X}})=2[1-{\rm NCF}(\textbf{\textit{X}})]. \label{eq:NSF}
\end{equation}
 Therefore, we can obtian the quadrupole moment ratio arising from the variable $f(\textbf{\textit{X}})$
 \begin{equation}
\frac{M_{\rm 2}(R)}{M_{\rm 0}(R)}=
\frac{\int_0^{2\pi} e^{-2i\varphi} {\rm NSF}(R,\varphi)~d{\varphi}}
{\int_0^{2\pi}  {\rm NSF}(R,\varphi)~d{\varphi}}  \label{eq:qudro}
\end{equation}
to reveal the spatial anisotropy, where $R$ is a radial separation and $\varphi$ the polar angle.

\subsection{Gradient Calculation of 2D images}\label{GMT}
 In the study of synchrotron gradient measurements, the Sobel operator\footnote{\url{http://en.volupedia.org/wiki/Sobel_operator}} will be used to compute an approximation of the gradient of the 2D image. This method adopts two 3x3 kernels that are convolved with the original 2D image to obtain approximations of the horizontal and vertical derivatives. And then, the recipe of sub-block averaging (\citealt{Yuen17}) for the gradient map obtained is used to determine the gradients of
of the subregion on the 2D image. In each subregion of interest, we use a Gaussian fitting method for the gradient angles to get an optimal direction characterized by the peak of the fitting.

\subsection{Alignment Measure}\label{AM}
For the magnetic field data cubes obtained by numerical simulation, we can know the intrinsic direction of the underlying magnetic field. Therefore, the magnetic field direction obtained by correlation function anisotropy and gradient techniques can be used to make a comparison with the inherent magnetic field direction. In the case of correlation function anisotropy, the major axis direction of the contour determine the magnetic field direction, whereas in the case of gradient technique, the direction of rotated $90^\circ$ gradients would identify the directions of the magnetic field.

We adopt a reduction factor to measure the correspondence between the measured magnetic field direction and the intrinsic magnetic field direction (see \citealt{GCL17})
\begin{equation}
AM=2\langle\cos^2\theta \rangle-1, \label{eq:AM}
\end{equation}
which is so called the alignment measure (AM), analogous to the Rayleigh reduction factor in dust alignment theory suggested by \cite{Greenberg68}. The parameter $\theta$ in Equation (\ref{eq:AM}) is the angle between the measured magnetic field direction and the intrinsic magnetic fields. $AM =\pm1$ represents a perfect alignment while $AM=0$ no alignment.

\section{Magnetic Turbulence Data Generation and Anisotropy Analysis}\label{3Ddata}
\subsection{Data Generation of Magnetic Turbulence}
\begin{table}
 \caption {Data cubes with numerical resolution of $512^3$ generated in the simulation of compressible MHD turbulence. $\delta B_{\rm rms}$ denotes the root mean square of random magnetic field, ${\langle B \rangle}$ the regular magnetic field and $\beta=2M_{\rm A}^2/M_{\rm s}^2$ the plasma parameter. }
 \centering
 \begin{tabular}{c c c cc}%
 \hline
Model & $M_{\rm A}$ & $M_{\rm s}$ & $\delta B_{\rm rm s} / \langle B \rangle$ & $\beta$\\ \hline 
run1 & 0.65  & 0.48 &  0.164 & 3.668  \\
run2& 0.70 & 0.87  &  0.579  &  1.295 \\
run3 & 0.55 & 4.46  & 0.467 & 0.030   \\
run4 & 0.50 & 9.92  & 0.465 & 0.005   \\ \hline 
\end{tabular}
\label{table:simdata}
\end{table}

The following equations are used to describe the interstellar magnetic turbulence environment where the synchrotron radiation is emitted
\begin{equation}
{\partial \rho }/{\partial t} + \nabla \cdot (\rho \textbf{\textit{v}})=0, \label{eq:11}
\end{equation}
\begin{equation}
\rho[\partial \textbf{\textit{v}}/{\partial t} + (\textbf{\textit{v}}\cdot \nabla) \textbf{\textit{v}}] +  \nabla p_g
        - \textbf{\textit{J}} \times \textbf{\textit{B}}/4\pi =\textbf{\textit{f}}, \label{eq:12}
\end{equation}
\begin{equation}
{\partial \textbf{\textit{B}}}/{\partial t} -\nabla \times (\textbf{\textit{v}} \times\textbf{\textit{B}})=0,\label{eq:13}
\end{equation}
\begin{equation}
\nabla \cdot \textbf{\textit{B}}=0. \label{eq:14}
\end{equation}
Here, $p_g=c_{\rm s}^2\rho$ is the gas pressure, $t$ the evolution time of the turbulent fluid, $\textbf{\textit{J}}=\nabla \times \textbf{\textit{B}}$ the current density, and $\textbf{\textit{f}}$ an external driving force. Additionally, an isothermal equation of state would close the above equations. 

Numerically, the third-order-accurate hybrid, essentially non-oscillatory code well checked is used to obtain data cubes reaching steady state. In our simulation, an external magnetic field with $B_0\simeq{\langle B \rangle}=1$ is set along the $x$-axis (horizontal) direction and the random turbulence is driven by a solenoidal driving force at the wavenumber $k=2.5$. The resulting 3D data cubes with numerical resolution of $512^3$, including the information of three 3D magnetic fields, three 3D velocities and one 3D density, are listed in Table \ref{table:simdata} where they are depicted by the Alfv\'enic and sonic Mach numbers. 

\subsection{Anisotropy Analysis of Magnetic Turbulence Data}\label{AAEdd}

\begin{figure*}[t]
\centering
\includegraphics[width=0.99\textwidth]{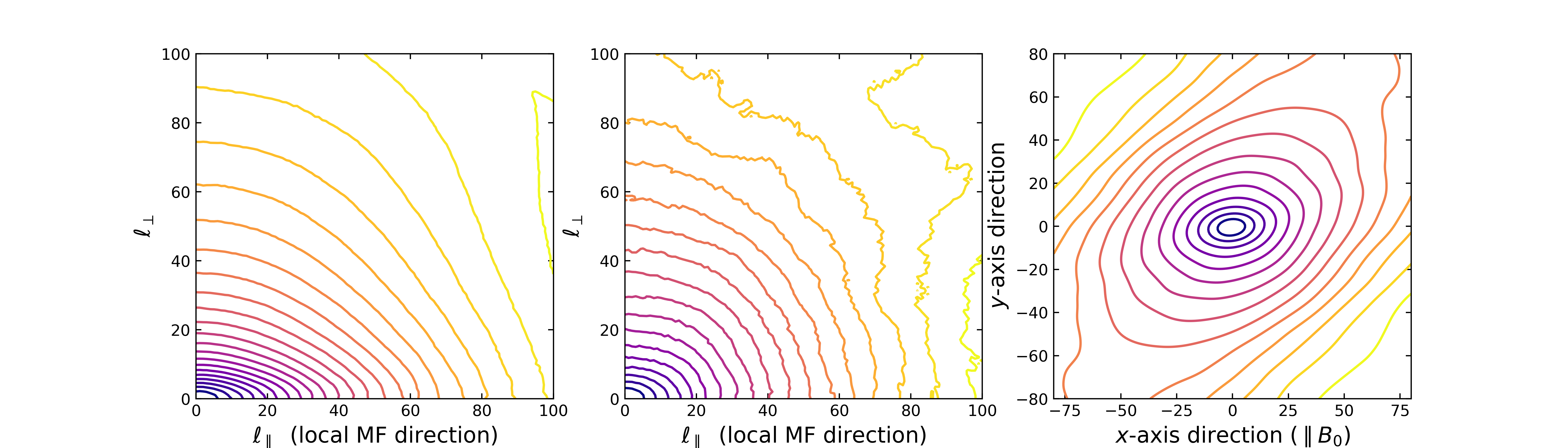}
\caption{Structures of scale-dependent anisotropic eddies in terms of the run3 magnetic field (MF) data listed in Table \ref{table:simdata}. The eddy anisotropies in 3D space are plotted in the left panel. The eddy anisotropies within the 100th 2D slice are respectively imaged along the major and minor axes of eddies (middle panel) and along the $x$ and $y$ axes (right panel).
} \label{fig:localmf}
\end{figure*}

Adopting a cylindrical coordinate system fixed on the eddies (following CV00), we obtain the local magnetic field by 
\begin{equation}
\textbf{\textit{B}}_l=[\textbf{\textit{B}}(\textbf{\textit{r}}_2) + \textbf{\textit{B}}(\textbf{\textit{r}}_1)    ]/2, \label{eq:localmf}
\end{equation}
where $\widehat{\ell_\parallel} =\textbf{\textit{B}}_l / | \textbf{\textit{B}}_l |$ is parallel to the major-axis direction of eddies. In the local reference frame of the magnetic field is written the corresponding second-order structure function as 
\begin{equation}
{\rm SF}(\ell_\parallel, \ell_\perp)=\left< |\textbf{\textit{B}}(\textbf{\textit{r}}_2) - \textbf{\textit{B}}(\textbf{\textit{r}}_1)|^2  \right>  \label{eq:localsf}
\end{equation}
with $\ell_\parallel=\widehat{\ell_\parallel}\cdot (\textbf{\textit{r}}_2-\textbf{\textit{r}}_1)$ and $\ell_\perp=|\widehat{\ell_\parallel} \times (\textbf{\textit{r}}_2-\textbf{\textit{r}}_1)  |$. 

 As a case in point, with the run3 magnetic field data cubes listed in Table \ref{table:simdata}, Figure \ref{fig:localmf} shows the structures of scale-dependent anisotropic eddies. The eddy anisotropies in 3D space are plotted in the left panel, while the eddy anisotropies within the 100th 2D slice are plotted along the major and minor axes of eddies (middle panel) and along the $x$ and $y$ axes (right panel), respectively. As seen in Figure \ref{fig:localmf}, the scale-dependent anisotropies of the eddies of MHD turbulence can indeed be revealed in the local reference frame, as predicted in GS95 and confirmed in CV00. The smaller the scale is, the more pronounced the degree of anisotropy is. In this figure is the result of the anisotropic structure of eddy within 2D slice, making a comparison with the synchrotron observations studied in the following. However, even for a plane-like (a slice) geometry, from the synchrotron observational data, we do not know a priori information about the magnetic field and eddy. As a result, we can study their properties only in the artificially selected ``local" coordinate system to observe the eddy structures, as done in the right panel of Figure \ref{fig:localmf}.

\section{Numerical results}\label{Res}
\subsection{Synthesis of Polarized Synchrotron Data Cubes}
\begin{figure*}[t]
\centering
\includegraphics[width=0.9\textwidth,bb=35 35 650 550]{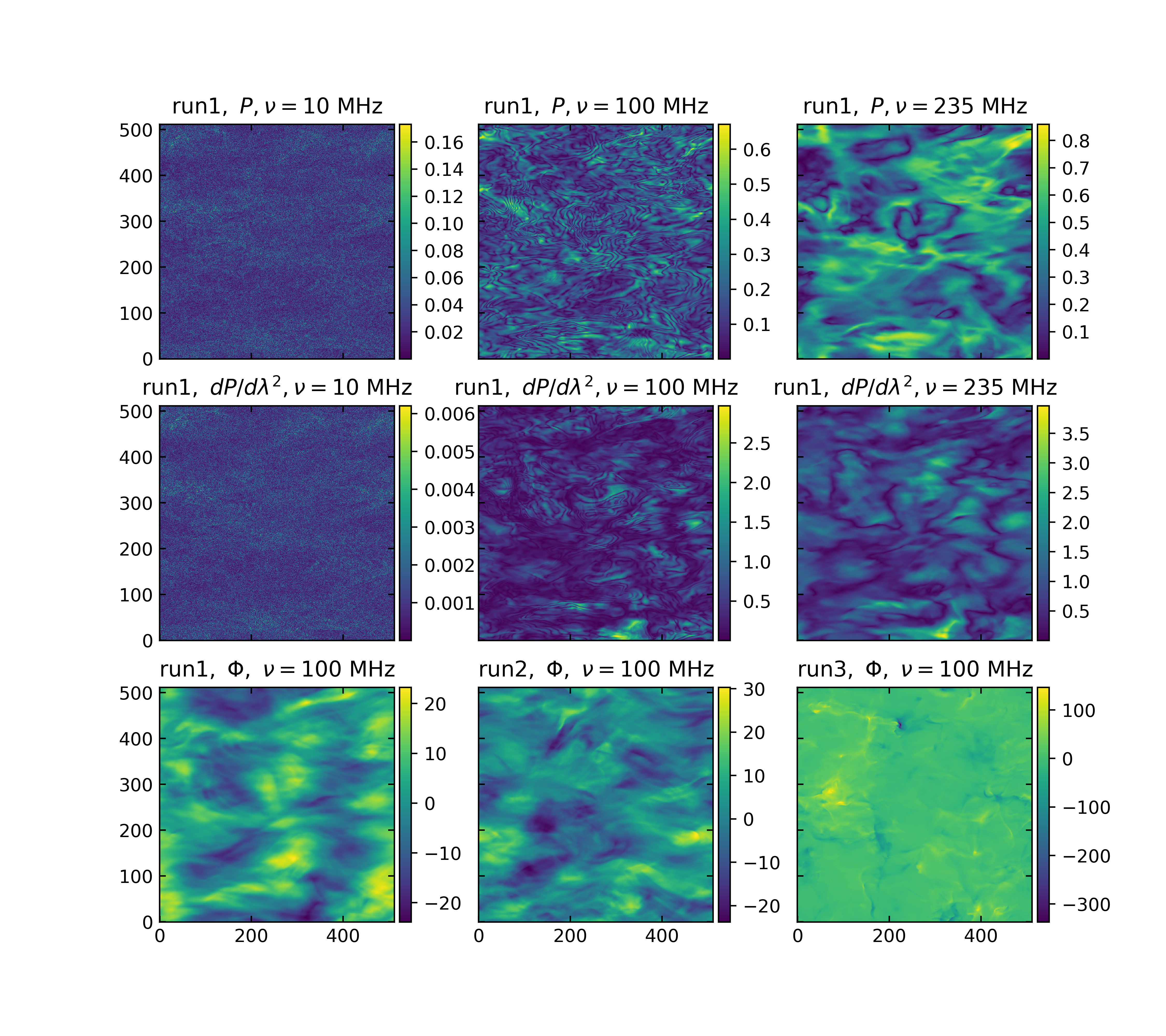}
\caption{Synchrotron polarization intensities (upper panels) and their derivatives with respect to the squared wavelength (middle panels) in units of mean synchrotron intensity, and Faraday rotation measure (lower panels) in units of rad $\rm m^{-2}$.
} \label{fig:maps}
\end{figure*}
For the sake of simplicity, we consider the scenario of the spatially coincident polarized synchrotron emission and Faraday rotation regions for the purpose of real observation simulating and the feasibility of various statistic technique testing, with the assumption of a spatial scale 1 kpc, the thermal electron density $0.01\ \rm cm^{-3}$, and the magnetic field strength 1 $\rm \mu G$. Being low-frequency-observation-motivated from Low Frequency Array for radio astronomy (LOFAR), we can synthesize the data cubes of Stokes parameters $I$, $Q$ and $U$ ranging from the frequency $\nu=10$ to 240 MHz in the bandwidth of $\Delta \nu=0.46$ MHz, on the basis of 3D simulation data listed in Table \ref{table:simdata}. This choice of the bandwidth can ensure a sufficiently small spatial interval in the line of sight direction to achieve the purpose of measuring the local magnetic field. 

Figure \ref{fig:maps} depicts the maps at individual frequency points so as to get a qualitative understanding of the synthesized observation data. The upper panels of Figure \ref{fig:maps} are the images of synchrotron polarization intensity $P=\sqrt{Q^2+U^2}$ while the middle panels correspond to the images of synchrotron polarization intensity derivative with regard to the squared wavelength $dP/d\lambda^2$. These images are computed for sub-Alfvénic and sub-sonic turbulence, using the run1 data of Table \ref{table:simdata}.  As seen, their structures are extended along the ($x$-axis) horizontal direction, for the mean magnetic field is set to this direction. We find that the precent of strong noise-like structure prevents our observation at the lowest frequency $10$ MHz. In general, these small-scale structure can be smoothed by a Gaussian filter technique. 

The lower panels of Figure \ref{fig:maps} are the images of the Faraday rotation measure, using data cubes from run1 to run3 of Table \ref{table:simdata}. The values of Faraday rotation measure for sub-Alfvénic and sub-sonic turbulence (left and middle lower panels) are smaller than those of sub-Alfvénic and super-sonic turbulence (lower right panel). That is caused by the occurrence of the high density (clump) regions arising form shock wave interactions in the case of super-sonic turbulence.

\subsection{Spectral Correlation Analysis of Synchrotron Polarization Derivative}\label{SCASPD}

\begin{figure*}[t]
\centering
\includegraphics[width=0.48\textwidth]{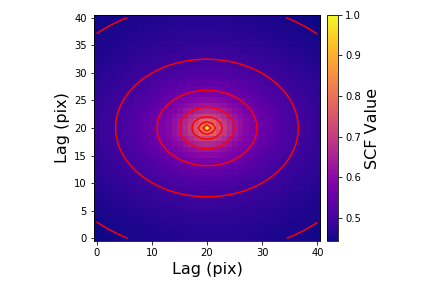}
\includegraphics[width=0.48\textwidth]{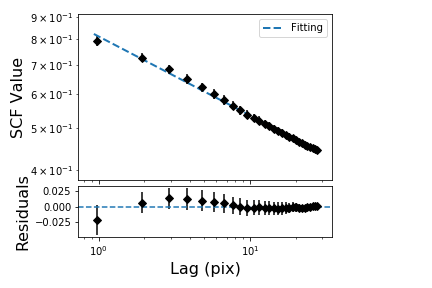}
\caption{Left panel: the image of SCF value of $dP/d\lambda^2$ and its fitting (solid line) from an elliptical power-law model, resulting in an index of $-0.191$. Right panels: an azimuthal average of the correlation surface (left panel) fitted by a linear least-squares method, with the power-law slope of $-0.182$, from which the inverse squared standard deviation is used as the weights. In the right lower panel is the residuals of the linear fitting. All calculations are based on the run3 listed in Table \ref{table:simdata}.
} \label{fig:SCF}
\end{figure*}
In this section, we adopt the SCF to explore the correlation of multi-frequency data cubes as a function of spatial lag separation. Numerical calculation is implemented by extracting the source code from a python package called TURBUSTAT (\citealt{Koch17}). With the synchrotron simulation data cubes generated by the run3 listed in Table \ref{table:simdata}, the resulting image of SCF of $dP/d\lambda^2$ is shown in left panel of Figure \ref{fig:SCF}, in which the solid contour line indicates the 2D fitting by an elliptical power-law model, obtaining an optimal index of $-0.191$. Interestingly, we see the solid contour line extending along the $x$ axis, which would reflect the mean magnetic field direction. Furthermore, an azimuthal average of the SCF image produces a 1D spectrum of the power-law slope of $-0.182$ (left upper panel), which is fitted by a linear least-squares method with the weights of the inverse squared standard deviation. We find that the 2D fitting gives slightly greater index than that of the 1D fitting. 

Similarly, the fitting results of other synchrotron simulation data are as follows: -0.194 (2D) and -0.181 (1D) for run1; -0.175 (2D) and -0.162 (1D) for run2; -0.179 (2D) and -0.172 (2D) for run4. All the above fittings, in the lag range from 1 to 40 pixels, demonstrate the correlation index around -0.2. On the other hand, we also study the possible correlation of SCF values of synchrotron polarization intensity $P$ but find a weak correlation, i.e., the fitting index less than -0.07 for four sets of synchrotron simulation data above. SCF of $dP/d\lambda^2$ reflects the information of the same projected mean magnetic field measured at different frequencies while SCF of $P$ reveals links between different local magnetic fields. Accordingly, we point out that the SCF analysis of $dP/d\lambda^2$ can be used to reveal the scale-dependent anisotropies of MHD turbulence. Similar to the studies of velocity channel maps (\citealt{Lazarian02}), it is expected that the frequency channel maps should disclose more information for the underlying magnetic field; this will be studied in detail in the future.

\subsection{Anisotropy Analysis of Synchrotron Polarization Derivative }
\begin{figure}[t]
\centering
\includegraphics[width=0.5\textwidth]{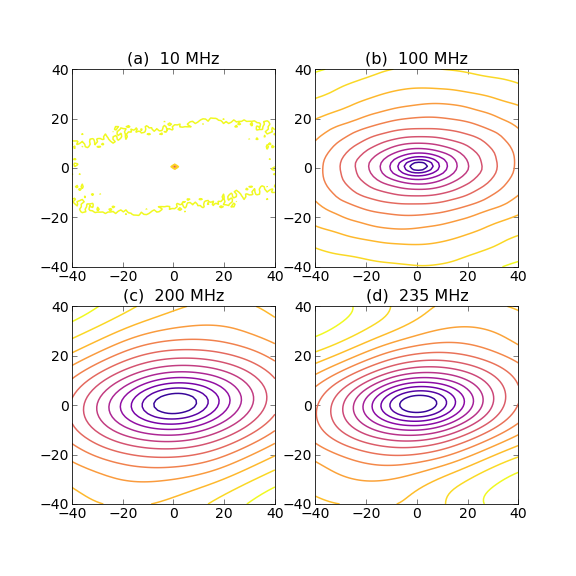}
\caption{Visualization of the structure function of $dP/d\lambda^2$, calculated by run1 listed in Table \ref{table:simdata} at different frequencies.
 } \label{fig:MFcontour}
\end{figure}

As mentioned in Section \ref{AAEdd}, since a priori local reference system cannot be fixed on the eddies for observational data, we select the Cartesian coordinate system on the plane of the sky.  Figure \ref{fig:MFcontour} plots the contour of the structure function of $dP/d\lambda^2$, which is calculated by the run1 of Table \ref{table:simdata} at frequencies 10 MHz ((a) panel), 100 MHz ((b) panel), 200 MHz ((c) panel) and 235 MHz ((d) panel). According to Equation (\ref{eq:el}), those frequencies are related to the spatial positions 1.4, 134.8, 570.7 and 790.2 pc (1 kpc extent length adopted along the line of sight), respectively, where the anisotropic structures of $dP/d\lambda^2$ qualitatively reflect those of the eddies. Exceptionally, no correlation for the 10 MHz simulation can be seen due to the influence of extremely strong numerical noise.

\begin{figure*}[t]
\centering
\includegraphics[width=0.95\textwidth]{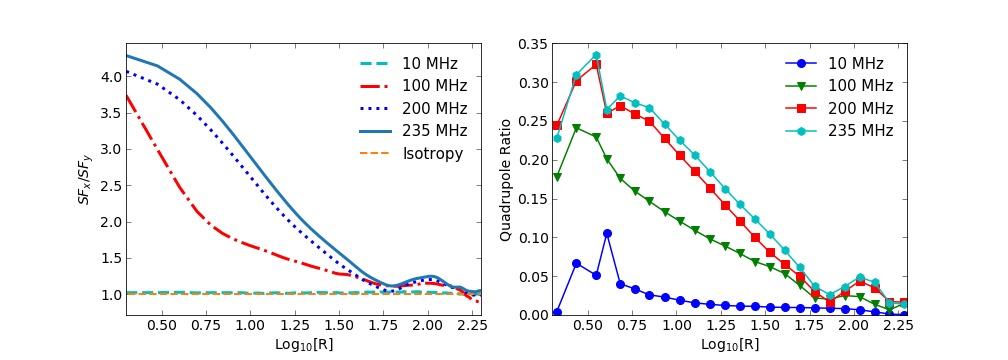}
\caption{ The ratio of the $x$-axis component to the $y$-axis component from the structure function of $dP/d\lambda^2$ (left panel) and the quadrupole ratio moduli of $dP/d\lambda^2$ (right panel) as a function of the radial separation, calculated by run1 listed in Table \ref{table:simdata} at different frequencies. The horizontal dashed line 
in the left panel represents isotropy. } \label{fig:QRM}
\end{figure*}

Quantitative analysis of the structure function of $dP/d\lambda^2$ is carried out in the left panel of Figure \ref{fig:QRM}, according to Equation (\ref{eq:SFxSFy}). As shown, the statistics at high frequencies present stronger anisotropy than those at low frequencies. The degree of the anisotropy of $dP/d\lambda^2$ decreases with the increase of the spatial scale, which reveals the scale-dependent anisotropy of MHD turbulence (GS95). Moreover, the quadrupole moment ratio (refer to Equation (\ref{eq:qudro})) modulus for $dP/d\lambda^2$, as a function of the spatial separation, is plotted in the right panel of Figure \ref{fig:QRM} at the corresponding frequencies. The quadrupole moment ratio moduli decrease with the increase of the spatial separation, which is generally consistent with the results from structure function of $dP/d\lambda^2$ (left panel). In particular, the quadrupole ratio modulus is more sensitive in the anisotropy analysis, as shown in the 10 MHz curves.

\subsection{Measurements of Magnetic Field Directions by Correlation Function Anisotropy}\label{MMFDCFA}

\begin{figure}[t]
\centering
\includegraphics[width=0.5\textwidth,bb=30 30 480 490]{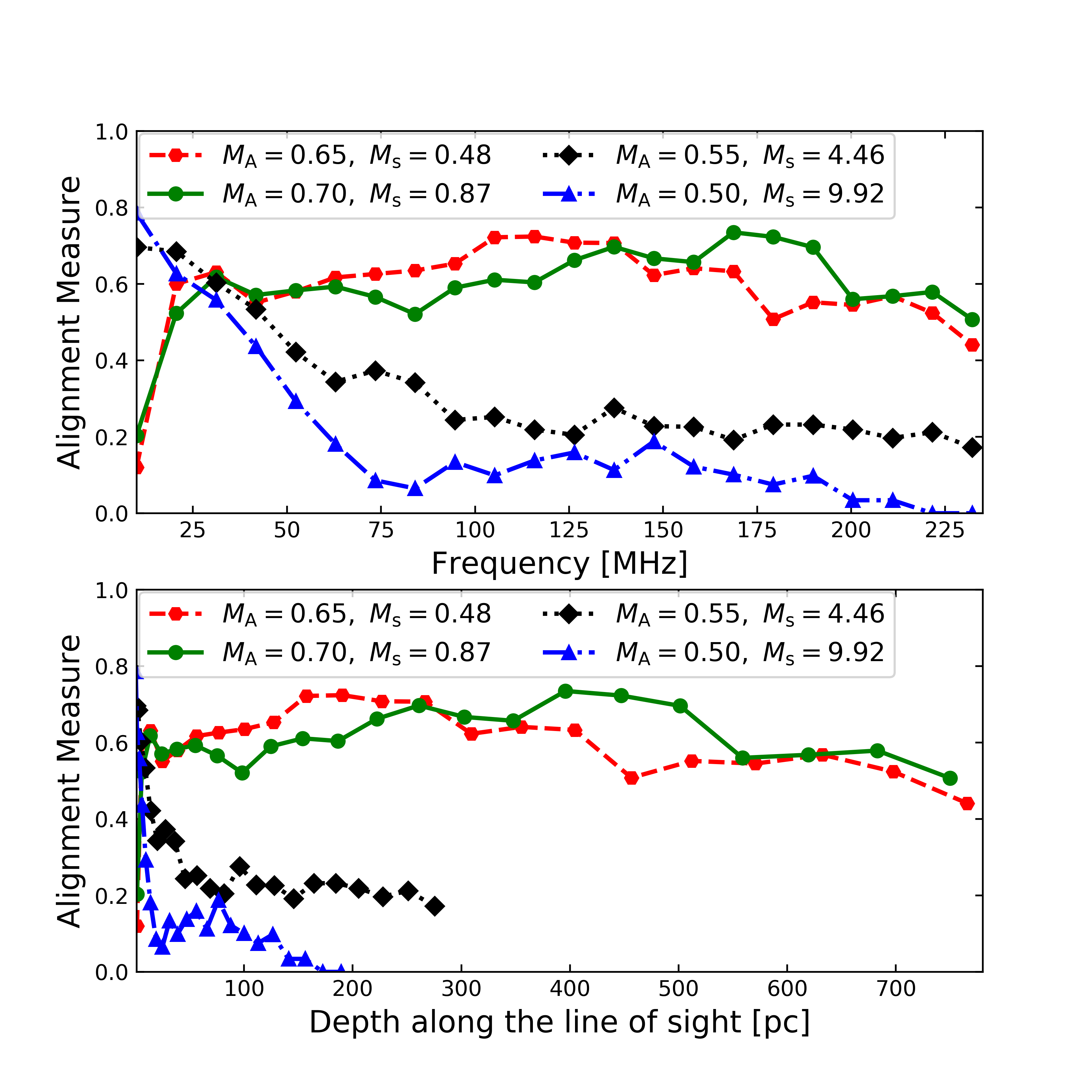}
\caption{Alignment measure between directions of the local magnetic fields and directions predicted by the CFA as a function of the frequency (upper panel) and of the spatial depth of Faraday rotation sampling (lower panel), on the basis of the data cubes listed in Table \ref{table:simdata}. 
} \label{fig:CFA}
\end{figure}
On the basis of the analysis of synchrotron polarization derivative in Section \ref{SynPD}, the correlation function anisotropy of $dP/d\lambda^2$ is able to reveal the magnetic turbulence anisotropy of the local spatial region. Measuring the major axis direction of the contour from the correlation function of $dP/d\lambda^2$, we could trace the local magnetic field direction. Similar to \cite{Yuen18} and \cite{Yang20}, we rotate the contour map to determine the orientation of the major axis. Compared with the earlier numerical implementation, which measured at single spatial point with the rotation step size 1 degree, this work makes two improvements, including the decrease of the step size to 0.5 degree and the average of the measurement direction of multiple spatial positions. 

The alignment measure, between intrinsic directions of the local magnetic fields and directions predicted by the CFA, is shown in Figure \ref{fig:CFA} as a function of radiation frequency (upper panel) and of the spatial depth along the line of sight (lower panel). In the case of the sub-Alfv\'enic and sub-sonic turbulence, i.e., $M_{\rm A} <1$ and $M_{\rm s} <1$, the CFA can trace the local magnetic field direction apart from the very-low-frequency part. This is because the presence of strong numerical noise affects the anisotropic distribution of the correlation function of $dP/d\lambda^2$, and the lack of anisotropy causes CFA not to work. When smoothing the small-scale noise structure in the low frequency range, we find that AM can be improved to some extent, but not shown in the paper. In the case of the sub-Alfv\'enic and super-sonic turbulence, i.e., $M_{\rm A} <1$ and $M_{\rm s} >1$, the adverse effect of numerical noise moves to a lower frequency part compared with the situation above. However, AM values decrease with the increase of frequency. In this regard, the formation of shock wave results in the strong density fluctuations. 
The density-dominated fluctuations should alter the fluctuations of random magnetic field, leading to the change of anisotropic structure of polarization derivative map, on which the CFA tracing is strongly dependent. 

As seen in Equation (\ref{eq:el}), the relationship between multiple frequencies can be mapped to the local spatial positions in the direction of the line of sight. In the lower panel of Figure \ref{fig:CFA} is AM vs. the corresponding spatial depth of Faraday rotation sampling. It can be seen that the sampling depth for sub-Alfv\'enic and sub-sonic turbulence is greater than that for sub-Alfv\'enic and sup-sonic turbulence. The reason is that the latter with high density fluctuations presents a stronger Faraday rotation than that of the former. It is understandable that strong Faraday rotation sampling depth is shallower. As a result, the CFA is effective for tracing the local magnetic field direction in the case of sub-Alfv\'enic and sub-sonic turbulence.

\subsection{Measurements of Magnetic Field Directions by Synchrotron Polarization Derivative Gradients}\label{MMFDSPDG}

\begin{figure}[t]
\centering
\includegraphics[width=0.5\textwidth,bb=30 30 480 490]{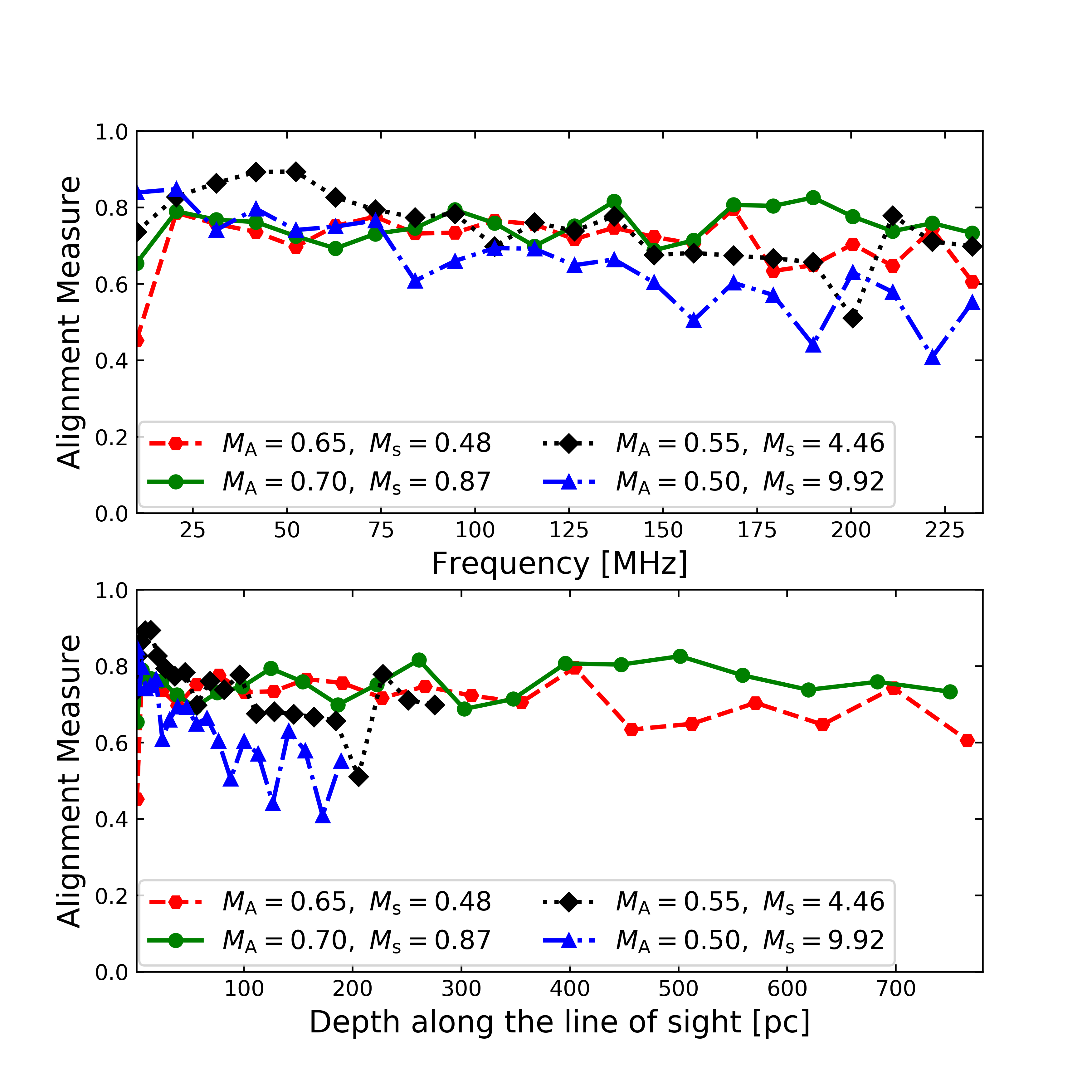}
\caption{Alignment measure between directions of the local magnetic fields and directions predicted by the SPDG as a function of the frequency (upper panel) and of the spatial depth of Faraday rotation sampling (lower panel), on the basis of data cubes listed in Table \ref{fig:SPDG}. 
} \label{fig:SPDG}
\end{figure}

Based on the gradient measurement method stated in Section \ref{GMT}, we first compute the gradients of $dP/d\lambda^2$ by the Sobel operator and divide the whole gradient map obtained into $8\times8$ sub-blocks, each of which has the numerical resolution of $64\times64$ pixels. We then carry out the Gaussian fitting for sub-block gradient map, the peak of which is considered as the optimal gradient orientation in each sub-block region. Finally, we average all AMs, alignment between the intrinsic magnetic field direction and the gradient one, to obtain an AM value at each frequency. 

The resulting AM is plotted in Figure \ref{fig:SPDG} as a function of the frequency (upper panel) and of the spatial depth of Faraday rotation sampling (lower panel), on the basis of four sets of data cubes listed in Table \ref{table:simdata}. In this figure, most simulations show that  AM remains almost unchanged as frequency increases, with an exception of the scenario in the super-sonic Mach ($M_{\rm s}\sim10$) turbulence where the AM decreases slightly. In particular, the present of numerical noise at approaching 10 MHz does not impede the tracing of magnetic field direction. It turns out that the synchrotron polarization derivative gradient (SPDG) technique is providing the ideal AM in the tracing of the local magnetic field. In comparison with the results provided in Figure \ref{fig:CFA} from the CFA measurement (Section \ref{MMFDCFA}), we find that the capability of the SPDG is greater than that of the CFA in the measurement of magnetic field directions, which is in agreement with the studies on velocity gradient techniques (\citealt{Yuen18,Yang20}). On the other hand, our previous studies demonstrated that the CFA of polarization intensities has the advantage of distinguishing compressible turbulence modes (LP12; \citealt{Herron16,Lee19,Wang20}). Therefore, we believe that the CFA and gradient analysis of synchrotron polarization derivative are complementary in the measurement of magnetic fields. 

\section{Discussion}
Statistic techniques of synchrotron polarization derivative is motivated by the anisotropic properties of strong MHD turbulence theory (GS95). The empirical evidence strongly supports the GS95 scale-dependent anisotropy $l_{\|}\propto l_{\perp}^{2/3}$ over its competitors. However, we did not discuss what theory can describe the MHD turbulence phenomenon correctly. Our focus lies in how to reveal the anisotropic properties of eddies and the direction tracing of the local magnetic field from synchrotron observations rather than confirm the accurate 2/3 scaling relation of GS95. The numerical results presented in this paper are independent of the particular turbulence phenomenology adopted. 

In Sections \ref{MMFDCFA} and \ref{MMFDSPDG}, no technical improvements were made in the AM values obtained, such as smoothing technique, optimizing fitting ways and improving sub-block averaging method as mentioned in \cite{Zhang19a}, through which the measurement level of the AM would be further enhanced. With strong external magnetic field set ($B_0>1$, accompanying with a strong mean magnetic field $\left<B\right>$) to get 3D data cubes in the previous studies of synchrotron gradients (LY18; \citealt{Zhang19a,Zhang19b}), the AM value should be greater to some extent because of its alignment between the projected mean magnetic fields and the 90-degree-rotated gradient directions. In this work, the reliability of the measurement results can be ensured when AM is greater than $\sim0.6$ since all simulations are run with the external magnetic field $B_0=1$. Besides, in the alignment measurement, the gradient technique is more dependent on the resolution of the sub-block than the CFA technique. Further studies are expected to employ higher resolution data in the future.

We would like to emphasize that magnetic field measurement techniques, such as synchrotron intensity fluctuations (LP12; \citealt{Herron16,Lazarian17}), synchrotron polarization intensity fluctuations (LP16; \citealt{Zhang16,Zhang18,Zhang19a,Lee16}; LY18) and synchrotron advanced diagnostics statistics (\citealt{Herron18a,Herron18b,Zhang19b}), have been used to investigate the anisotropy caused by the mean magnetic field, the direction of the projected mean magnetic field and the power spectral distributions. Among them, the SPDG technique is proposed to reveal the \textit{locally projected mean}  magnetic field direction within the slices of position-position-position (PPP) data cubes (LY18 and \citealt{Zhang19a}). In order to reveal the local magnetic field properties, this work requires a sufficiently narrow frequency bandwidth, which will be ensured to correspond spatially to a single PPP slice, with no projected effect involved. From an observational point of view, our setting a frequency resolution of 0.64 MHz in this paper will not be a problem. 

The well-known Faraday tomography method proposed by \cite{Burn66} defines the Faraday dispersion function as a Fourier transform of the polarization surface brightness with respect to the squared wavelength $\lambda^2$, aiming at obtaining the information of intrinsic polarization intensity considered as a function of Faraday depth (\citealt{Brentjens05}). However, the current statistic measurements related to synchrotron polarization fluctuations define the complex polarized vector as a function of the squared wavelength $\lambda^2$, and avoid a Fourier transform which would mathematically result in the disorder between the Faraday depth and its transform function. A comparison between Faraday Tomography and synchrotron gradient measurement techniques is made to distinguish the effectiveness of the two techniques (\citealt{Ho19}). Our techniques can be seen as a complement to the Faraday tomography. It is expected that a synergetic application of these techniques will provide a larger amount of valuable information on the magnetic field in the future.

\section{Summary}
With the data cubes generated in numerical simulations, we have investigated in this paper the scale-dependent anisotropy of eddies and the local magnetic field direction by the statistics of synchrotron polarization derivative with respect to the squared wavelength, involving the structure function ratio, quadrupole ratio modulus, spectral correlation function, correlation function anisotropy analysis and gradient measurement. The resultant findings are listed as follows.

\begin{enumerate}[wide, labelwidth=!, labelindent=1pt]
\item Statistic analysis of $dP/d\lambda^2$ reveals the scale-dependent anisotropy of underlying MHD turbulence, that is, the smaller the spatial scale is, the more significant the degree of the anisotropy presents. In addition, the degree of the anisotropy revealed enhances with the increase of the radiation frequency . 

\item SCF analysis of $dP/d\lambda^2$ can be used to explore the scale-dependent anisotropy of magnetic turbulence. CFA of $dP/d\lambda^2$ has a capability of tracing the underlying local magnetic field direction in the case of the sub-Alfv\'enic and sub-sonic turbulence. 

\item  Gradient techniques of $dP/d\lambda^2$ work well in the measurement of the local magnetic field direction for sub-Alfv\'enic, sub- and super-sonic turbulence regimes. 

\item Being extremely low-frequency-regime-oriented, the synergy of diverse techniques of $dP/d\lambda^2$ paves a way for the application of the LOFAR data cubes to the study of the MHD turbulence. 
\end{enumerate}

\acknowledgments
We thank the anonymous referee for his or her valuable comments improving our manuscript. J.F.Z. thanks the supports from the National Natural Science Foundation of China (grant Nos. 11973035 and 11703020), the Hunan Provincial Natural Science Foundation (grant No. 2018JJ3484) and the Guizhou Provincial Key Laboratory of Radio Astronomy and Data Processing (grant No. KF201803). K.H. acknowledges the support of the National Natural Science Foundation of China (grant Nos. U1931115, U1731110 and U1731106). J.C. is thankful for the support from National R\&D Program through the National Research Foundation of Korea Grants funded by the Korean Government (NRF-2016R1A5A1013277 and NRF- 2016R1D1A1B02015014). A.L. acknowledges the support of NSF grant AST 1715754.


\begin{thebibliography}{}
\providecommand\natexlab[1]{#1}
\providecommand\JournalTitle[1]{#1}
\bibitem[Armstrong et al. (1995)]{Armstrong95} Armstrong, J.~W., Rickett, B.~J., \& Spangler, S.~R.\ 1995,
\href{http://adsabs.harvard.edu/abs/1995ApJ...443..209A}{\JournalTitle{\apj}, 443, 209} 

\bibitem[Beresnyak \& Lazarian(2019)]{BL19} Beresnyak, A., \& Lazarian, A.\ 2019, Turbulence in Magnetohydrodynamics, 
\href{https://ui.adsabs.harvard.edu/abs/2019tuma.book.....B/}{\JournalTitle{tuma.book}, Berlin/Boston}

\bibitem[Biskamp (2003)]{Biskamp03}Biskamp, D. 2003, in Magnetohydrodynamic Turbulence, ed. D. Biskamp
(Cambridge: Cambridge Univ. Press), 1

\bibitem[Burkhart et al. (2012)]{Burkhart12} Burkhart, B., Lazarian, A., \& Gaensler, B.~M.\ 2012, 
\href{http://dx.doi.org/10.1088/0004-637X/749/2/145}{\JournalTitle{\apj}, 749, 145}

\bibitem[Burn (1966)]{Burn66}Burn, B. J. 1966, 
\href{http://dx.doi.org/10.1093/mnras/133.1.67}{\JournalTitle{MNRAS}, 133, 67}

\bibitem[Brentjens \& de Bruyn (2005)]{Brentjens05}Brentjens, M., \& de Bruyn, A. 2005,
\href{http://dx.doi.org/10.1051/0004-6361:20052990}{\JournalTitle{A\&A}, 441, 1217}

\bibitem[Chepurnov \& Lazarian (2010)]{Chepurnov10} Chepurnov, A., \& Lazarian, A.\ 2010,
\href{http://adsabs.harvard.edu/abs/2010ApJ...710..853C}{\JournalTitle{\apj}, 710, 853}

 \bibitem[Cho \& Lazarian(2010)]{Cho10}Cho, J., \& Lazarian, A. 2010, ApJ, 720, 1181
 
\bibitem[Cho \& Vishniac (2000)]{ChoV00} Cho, J., \& Vishniac E.~T.\ 2000, 
\href{http://adsabs.harvard.edu/abs/2000ApJ...539..273C} {\JournalTitle{ApJ},  539, 273} 

\bibitem[{Cho \& Lazarian (2002)}]{Cho02PRL} Cho, J., \& Lazarian, A. 2002,
\href{http://dx.doi.org/10.1103/PhysRevLett.88.245001}{\JournalTitle{PhRvL}, 88, 245001}

\bibitem[Cho et al. (2002)]{ChoLV02} Cho, J., Lazarian, A., \& Vishniac E.~T.\ 2002,
\href{http://adsabs.harvard.edu/abs/2002ApJ...564..291C} {\JournalTitle{ApJ}, 564, 291}

\bibitem[Koch et al. (2017)]{Koch17}Koch, E. K., Caleb., G. W., Offner, W., Loeppky, J. L., Rosolowsky, E. W. 2017, MNRAS, 471, 1506

\bibitem[Ferri{\`e}re(2016)]{Ferriere16} Ferri{\`e}re, K.\ 2016, 
\href{https://ui.adsabs.harvard.edu/abs/2016JPhCS.767a2006F} {\JournalTitle{J. Phys. Conf. Ser.}, 767, 012006} 

\bibitem[Gaensler et al. (2011)]{Gaensler11} Gaensler, B.~M., Haverkorn, M., Burkhart, B., et al.\ 2011,
\href{http://dx.doi.org/10.1038/nature10446} {\JournalTitle{\nat}, 478, 214 }

\bibitem[Galtier et al. (2000)]{Galtier00} Galtier S., Nazarenko S.~V., Newell A.~C., \& Pouquet A.\ 2000,
\href{http://adsabs.harvard.edu/abs/2000JPlPh..63..447G} {\JournalTitle{JPlPh}, 63, 447} 

\bibitem[{Getmantsev (1959)}]{Getm59}Getmantsev, G. G. 1959, Sov. Astron. Journ., 36, 422

\bibitem[Ginzburg \& Syrovatskii (1965)]{Ginzburg65}Ginzburg, V.~L., \& Syrovatskii, S.~I.\ 1965, \araa, 3, 297 

\bibitem[{Goldreich \& Sridhar (1995)}]{Gold95}Goldreich, P.~ \& Sridhar, S. 1995,
\href{http://dx.doi.org/10.1086/174600}{\JournalTitle{ApJ}, 438, 763}

\bibitem[Gonz{\'a}lez-Casanova \& Lazarian (2017)]{GCL17} Gonz{\'a}lez-Casanova D.~F., Lazarian A. 2017, ApJ, 835, 41

\bibitem[Guo et al. (2017)]{GuoMW17} Guo, X., Mao, J., \& Wang, J. 2017, ApJ, 843, 23

\bibitem[Greenberg (1968)]{Greenberg68} Greenberg, J. M. 1968, in Nebulae and Interstellar Matter, ed. B. M. Middlehurst \& L. H. Aller (Chicago, IL: Univ. Chicago Press), \href{http://adsabs.harvard.edu/abs/1968nim..book.....M }{11}

\bibitem[Herron et al. (2016)]{Herron16} Herron, C.~A., Burkhart, B., Lazarian, A., Gaensler, B.~M., \& McClure-Griffiths, N.~M.\ 2016, 
\href{http://dx.doi.org/10.3847/0004-637X/822/1/13}{\JournalTitle{\apj}, 822, 13}

\bibitem[Herron et al. (2018a)]{Herron18a} Herron, C.~A., Burkhart, B., Gaensler, B.~M., et al.\ 2018a,
\href{http://dx.doi.org/10.3847/1538-4357/aaafd0} {\JournalTitle{\apj}, 855, 29} 

\bibitem[Herron et al. (2018b)]{Herron18b} Herron, C.~A., Gaensler, B.~M., Lewis, G.~F., \& McClure-Griffiths, N.~M.\ 2018b, 
 \href{http://dx.doi.org/10.3847/1538-4357/aaa002}{\JournalTitle{\apj}, 853, 9 }
 
 \bibitem[Ho et al. (2019)]{Ho19}Ho, K. W., Yuen, K. H., Leung, P. K., Lazarian, A. 2019, ApJ, 887, 258
 
 \bibitem[Hockney (1968)]{Hock68}Hockney, R. W. 1968, PhFl, 11, 1381
 
 \bibitem[Junklewitz \& Enßlin (2011)]{Junk11}Junklewitz, H., \& Enßlin, T. A. 2011, A\&A, 530, A88

\bibitem[Kowal \& Lazarian (2010)]{Kowal10} Kowal, G., \& Lazarian, A.\ 2010, \apj, 720, 742

\bibitem[Lazarian (2009)]{L09}Lazarian, A. 2009, Space Sci. Rev., 143, 357

\bibitem[Lazarian \& Shutenkov (1990)]{L90}Lazarian, A. L., \& Shutenkov, V. P. 1990, SvAL, 16, 297

\bibitem[Lazarian \& Vishniac (1999)]{Lazarian99} Lazarian, A., \& Vishniac, E.~T. 1999,
  \href{http://dx.doi.org/10.1086/307233}{\JournalTitle{ApJ}, 517, 700}
  
\bibitem[{{Lazarian} \& {Pogosyan} (2012)}]{LP12} {Lazarian}, A., \& {Pogosyan}, D. 2012,
  \href{http://dx.doi.org/10.1088/0004-637X/747/1/5}{\JournalTitle{\apj}, 747, 5}
  
\bibitem[{{Lazarian} \& {Pogosyan} (2016)}]{LP16}  {Lazarian}, A., \& {Pogosyan}, D.\ 2016,
 \href{http://dx.doi.org/10.3847/0004-637X/818/2/178}{\JournalTitle{\apj}, 818, 178}

 \bibitem[Lazarian et al. (2017)]{Lazarian17} Lazarian, A., Yuen, K.~H., Lee, H., \& Cho, J.\ 2017, 
\href{http://dx.doi.org/10.3847/1538-4357/aa74c6}{\JournalTitle{\apjl}, 842, 30 }

\bibitem[Lazarian \& Yuen (2018)]{LY18} Lazarian, A., \& Yuen, K.~H.\ 2018,
\href{http://dx.doi.org/10.3847/1538-4357/aad3ca}{\JournalTitle{\apj}, 865, 59}

\bibitem[Lazarian et al. (2020)]{Lazarian20}Lazarian, A., Gregory, L. E., Jafari, A., et al. 2020, Phys. Plasmas, 27, 012305

\bibitem[Lazarian (2006)]{Lazarian06} Lazarian, A.\ 2006,
 \href{http://dx.doi.org/10.1086/505796}{\JournalTitle{\apjl}, 645, L25 }
 
 \bibitem[Lazarian et al. (2002)]{Lazarian02}Lazarian, A., Pogosyan, D., \& Esquivel, A. 2002, ASPC, 276, 182L
 
 \bibitem[Lee et al. (2016)]{Lee16} Lee, H., Lazarian, A., \& Cho, J.\ 2016, 
\href{http://dx.doi.org/10.3847/0004-637X/831/1/77}{\JournalTitle{\apj}, 831, 77 }

\bibitem[Lee et al. (2019)]{Lee19} Lee, H., Cho, J., \& Lazarian, A.\ 2019, \apj, 877, 108

\bibitem[Lithwick \& Goldreich(2001)]{Lithwick01} Lithwick Y., \& Goldreich P.\ 2001,
\href{http://adsabs.harvard.edu/abs/2001ApJ...562..279L}{\JournalTitle{ApJ}, 562, 279} 

\bibitem[Mac Low \& Klessen (2004)]{2004RvMP...76..125M} Mac Low, M.-M., \& Klessen, R. S.\ 2004, RvMP, 76, 125

\bibitem[Mao \& Wang (2018)]{Mao18a}Mao, J., \& Wang, J. 2018, ApJ, 854, 51

\bibitem[Mao et al. (2018)]{Mao18b}Mao, J., Covino, S., \& Wang, J. 2018, ApJ, 860, 153

\bibitem[Mao \& Wang (2013)]{Mao13} Mao, J., \& Wang, J. 2013, ApJ, 776, 17

\bibitem[McKee \& Ostriker (2007)]{2007ARA&A..45..565M} McKee, C.~F., \& Ostriker, E.~C.\ 2007, \araa, 45, 565

\bibitem[Narayan \& Medvedev (2001)]{2001ApJ...562L.129N} Narayan, R., \&  Medvedev, M.~V.\ 2001, \apjl, 562, L129 

\bibitem[Padoan et al. (2003)]{Padoan03}Padoan, P., Goodman, A.~A., Juvela, M. 2003, ApJ, 588, 881
       
\bibitem[Wang et al. (2020)]{Wang20}Wang, R.-Y., Zhang, J.-F. \& Xiang, F.-Y. 2020, ApJ, 890, 70

\bibitem[Waelkens et al.(2009)]{Wael09} Waelkens, A.~H., Schekochihin, A.~A., \& En{\ss}lin, T.~A.\ 2009,
\href{http://adsabs.harvard.edu/abs/2009MNRAS.398.1970W} {\JournalTitle{\mnras}, 398, 1970} 

\bibitem[Xu \& Zhang (2016)]{Xu16}Xu, S., \& Zhang, B. 2016, ApJ, 824, 113

\bibitem[Yang et al. (2020)]{Yang20} Yang, B., Zhang, J.-F., Lazarian, A., \& De Medeiros, J. R. 2020, MNRAS, submitted

\bibitem[Yuen et al. (2018)]{Yuen18} Yuen, K. H., Chen, J., Hu, Y., et al. 2018, ApJ, 865, 54

\bibitem[Yuen \& Lazarian (2017)]{Yuen17}Yuen, K. H., \& Lazarian, A. 2017, ApJL, 837, L24

\bibitem[Yan \& Lazarian (2008)]{Yan08} Yan, H., \& Lazarian, A.\ 2008, \apj, 673, 942

\bibitem[Zhang et al. (2017)]{Zhang17} Zhang, J.-F., Xiang, F.-Y \& Lu, J.-F. 2017, ApJ, 836, 72

\bibitem[Zhang et al. (2019a)]{Zhang19a} Zhang, J.-F., Lazarian, A., Ho, K.~W., et al.\ 2019a, 
\href{https://ui.adsabs.harvard.edu/abs/2019MNRAS.486.4813Z}{\JournalTitle{\mnras}, 486, 4813}
 
\bibitem[Zhang et al. (2018)]{Zhang18} Zhang, J.-F., Lazarian, A., \& Xiang, F.-Y.\ 2018,
\href{http://dx.doi.org/10.3847/1538-4357/aad182}{\JournalTitle{\apj}, 863, 197 }

\bibitem[Zhang et al. (2016)]{Zhang16} Zhang, J.-F., Lazarian, A., Lee, H., \& Cho, J.\ 2016, 
\href{http://dx.doi.org/10.3847/0004-637X/825/2/154}{\JournalTitle{\apj}, 825, 154}

\bibitem[Zhang et al. (2019b)]{Zhang19b} Zhang, J.-F., Liu, Q., \& Lazarian, A. 2019b, ApJ, 886, 63

\end{thebibliography}
\end{document}